\documentclass[aps,prl,twocolumn,groupedaddress,showpacs]{revtex4}

\usepackage{graphicx}

\begin{document}


\title{Direct evidence for ferromagnetic spin polarization in gold nanoparticles}


\author{Y. Yamamoto, T. Miura, T. Teranishi, M. Miyake and H. Hori}
\affiliation{School of Materials Science, Japan Advanced Institute of Science and Technology (JAIST), 1-1 Asahidai, Tatsunokuchi, Ishikawa, 923-1292, Japan}
\author{M. Suzuki, N. Kawamura, H. Miyagawa, T. Nakamura and K. Kobayashi}
\affiliation{Japan Synchrotron Radiation Research Institute (JASRI / SPring-8), 1-1-1 Kouto, Mikazuki, Sayou 679-5198, Japan}

\date{\today}

\begin{abstract}
 We report the first direct observation of ferromagnetic spin polarization of Au nanoparticles with a mean diameter of 1.9 nm using X-ray magnetic circular dichroism (XMCD). Owing to the element selectivity of XMCD, only the gold magnetization is explored. Magnetization of gold atoms estimated by XMCD shows a good agreement with the results obtained by conventional magnetometry. This result is evidence of intrinsic spin polarization in nano-sized gold.
\end{abstract}


\maketitle

Metal nanoparticles or clusters have recently become of great interest to physicists, chemists and engineers with regards their practical applications, such as high density storage media \cite{Sun} and next generation devices like single electron transistors \cite{Andres}, as well as for their fundamental physics \cite{Volokitin}. When the size of the particles or clusters is sufficiently small, degenerate energy levels split into discrete levels and the Fermi wavelength of the electrons becomes comparable to the size of system. Consequently, the physical properties of these particles are expected to be quite different from the bulk properties. However, due to the difficulty of preparing samples with well-defined diameters, for a long time it has been difficult to observe the difference in properties experimentally. In the past two decades, a new class of materials, metal nanoclusters stabilized by chemical materials, have been extensively studied \cite{Schmid1,Jongh}. In particular, noble metal nanoclusters have been synthesized using various stabilizer molecules, \textit{e.g.} Au$_{55}$(PPh$_{3})_{12}$Cl$_{6}$, Pd$_{561}$phen$_{36}$O$_{200\pm 20}$,$_{ }$etc \cite{Schmid2}. 
Recently, a size adjustment method for noble metal (Pd, Pt, Au) nanoparticles embedded in linear alkyl chain polymers has been developed \cite{Teranishi1,Teranishi2}. This method ensures the generation of mono-dispersed metal nanoparticles with arbitrary diameter. A wide range of particle diameters can be achieved and the variance in the diameter is very small. The ability of this method to adjust the size of the clusters allows a systematic investigation of the physical properties of metal clusters in the nanometer size range.

Hori \textit{et al.} have reported the macroscopic magnetic properties of polymer stabilized metal nanoparticles \cite{Hori}. They observed superparamagnetic behavior in Pd and Au nanoparticles embedded in poly($N$-vinyl-2-pyrrolidone) (PVP) at low temperatures. This fact means that the individual Pd or Au particles possibly have ferromagnetic moments, although Pd shows paramagnetism and Au shows diamagnetism in the bulk states. However, no clear evidence has been provided that the ferromagnetism of Au particles is intrinsic, because their SQUID magnetization data may include magnetization signals originating from trivial sources other than the nanoparticles, such as magnetic impurities or unpaired electrons in the protective polymer. The effects from these sources cannot be removed using conventional magnetometry techniques. An experimental technique that allows the selective observation of the magnetization of Au nanoparticles should be used.

In this paper, we present direct evidence of the intrinsic ferromagnetism of Au nanoparticles by means of element-specific magnetization (ESM) measurements based on the X-ray magnetic circular dichroism (XMCD) technique \cite{Schutz2}. The XMCD technique allows the detection of the magnetic moments of a particular element through sensitivity to the difference between the up- and down-spin densities around the Fermi level. This element selectivity is the most important advantage of this technique over conventional magnetometry and is essential in the present study for extracting the magnetization of the Au nanoparticles. Since the first observation of XMCD in Fe \cite{Schutz}, the magnetism of nanostructured materials \cite{Gambardella}, thin films \cite{Weller} and strongly correlated systems \cite{Kucela} have been widely investigated using this method.

The XMCD spectra were recorded using a highly sensitive spectrometer \cite{Suzuki} installed at BL39XU of SPring-8 synchrotron radiation facilities. External magnetic fields up to 10 T were applied along the X-ray beam direction using a split-type superconducting magnet. The experimental resolution was high enough to detect XMCD signals of 10$^{-5}$ parts of the spin-averaged X-ray absorption coefficients. This high sensitivity is achieved through the helicity modulation technique \cite{Suzuki} based on lock-in detection, and the high brilliance of the third generation synchrotron radiation source. We emphasize that the modulation technique was crucial for the detection of very small ferromagnetism in Au nanoparticles because this technique dramatically improves the signal/noise ratio of the XMCD signal.

\begin{figure}[tbp]
\includegraphics[width=7cm]{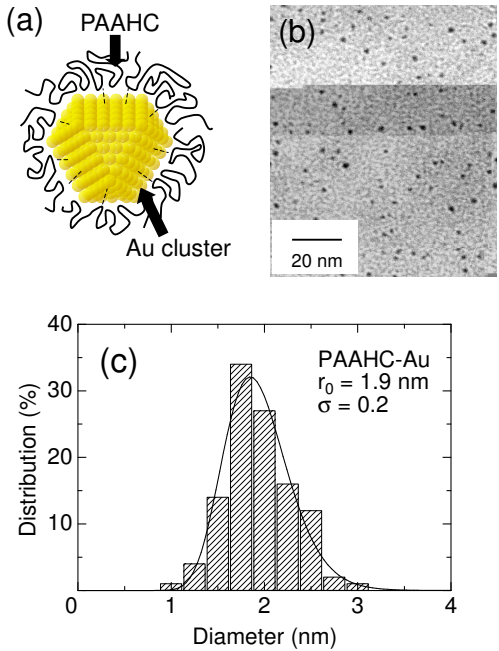}
\caption{\label{TEM} (a) Schematic illustration of the PAAHC-Au nanoparticles. (b) TEM image of Au nanoparticles protected by PAAHC. Mean diameter of the sample is 1.9 nm. Scale bar is shown in the image. (c) Size distributions of PAAHC-Au nanoparticles. The solid line represents the fitting curve assuming the log-normal function $f(r)=\frac{1}{\sqrt {2\pi } r\sigma }\exp \left( {-\frac{(\ln r-r_0 )^2}{2\sigma ^2}} \right)$ where $r$ is the particle diameter. The fitting results of parameters are the mean particle diameter $r_{0}$ = 1.9 nm and the standard deviation\textit{${\rm g}\sigma $} = 0.2 nm.}
\end{figure}

In the present work, Au nanoparticles protected by polyallyl amine hydrochloride (PAAHC) (abbreviated as PAAHC-Au) are used as samples, prepared by a method described in Ref.\cite{Teranishi1}. Each nanoparticle (cluster of Au metal) is surrounded by the matrix polymer, which prevents further aggregation and oxidation of the Au nanoparticles (Fig. 1(a)). The Au nanoparticles show a cuboctahedron shape with an \textit{fcc} structure. The nanoparticles are stabilized by the adsorption of the amide functional group of PAAHC and a hydrophobic interaction between the surface of the nanoparticles and the train part of the linear alkyl chain (dotted lines in Fig. 1(a)). The tail part of the polymer forms loops and shows folding with some conformation (thick solid line). The samples are stable in air for several months. The distribution of particle diameter was determined from a transmission electron microscope image (TEM, Hitachi H-7100). The mean diameter and standard deviation of the particles were 1.9 nm (212 atoms per cluster) and 0.2 nm respectively, measured by counting 200 particles from the TEM image (Fig. 1(b), (c)). The crystal structure of the nanoparticles was identified to be an \textit{fcc} structure by X-ray powder diffraction (XRD, MAC Science) using Cu-K$\alpha $ radiation. Magnetization and susceptibility measurements were performed using a commercial superconducting quantum interference device (SQUID) magnetometer (Quantum Design, MPMS-XL) in the temperature range between 1.8 and 300 K and a magnetic field up to 7 T.

\begin{figure}[tbp]
\includegraphics[width=7cm]{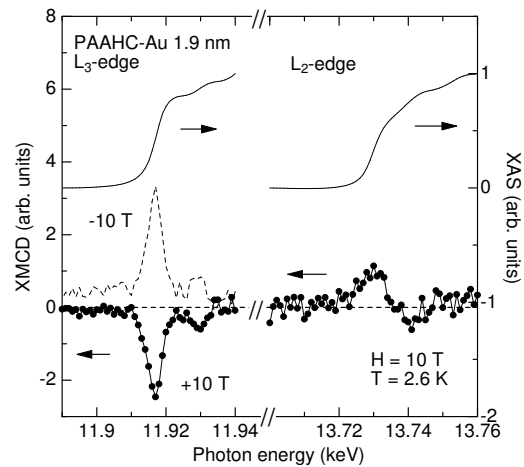}
\caption{\label{spectra} XMCD and XAS spectra at the Au $L_{3}$- and $L_{2}$-edge with an applied magnetic field of 10 T. The XMCD and XAS spectra include a same scale factor so that the height of the edge jumps of the XAS spectra (solid lines) are unity. Solid lines with closed circle symbols represent an XMCD spectrum with an applied magnetic field of 10 T, while the dotted line indicates an XMCD spectrum in a magnetic field applied in the opposite direction, implying the observed signal is not an artificial effect.}
\end{figure}

Figure 2 shows X-ray absorption spectroscopy (XAS) and XMCD spectra of Au at the $L_{3}$- (2$p_{3/2} \rightarrow $ 5$d_{5/2}$, 6$s_{1/2}$ dipole allowed transitions) and $L_{2}$-edge (2$p_{1/2} \rightarrow $ 5$d_{3/2}$, 6$s_{1/2})$ at 2.8 K in an applied magnetic field of 10 T. A negative XMCD signal was clearly observed at the $L_{3}$-edge (11.917 keV), whereas the XMCD signal at the $L_{2}$-edge (13.730 keV) was positive. The XMCD amplitude is of the order of 10$^{-4}$ of the XAS step height. Small negative peak structures were observed at the high-energy side of the main peaks, at 11.930 and 13.741 keV. As shown by the dotted line in Fig. 2, the sign of the XMCD signal reversed when the magnetic field direction was changed. This result verifies that the observed signal is truly of magnetic origin and does not arise from any artificial effects. To the best of our knowledge, this is the first observation of an evident XMCD signal arising from Au. 

The XMCD peak height at the $L_{2}$-edge is less than half that at the $L_{3}$-edge. This asymmetry indicates that the Au 5$d$ electrons have considerable orbital moment. Using the sum rules \cite{Thole,Carra}, the ratio of the orbital to the spin magnetic moments was determined to be $\langle\mu _{L}\rangle$/$\langle\mu _{S}\rangle$ = 0.145, which is larger than that in orbital-quenched 3$d$ transition metals and is comparable to the reported values for Pd and Pt systems \cite{Vogel,Wilhelm}. In determining $\langle\mu _{L}\rangle$/$\langle\mu _{S}\rangle$, the value of the hole number in the Au valence is not required and the magnetic dipole term $\langle T_{z}\rangle$ was neglected. In principle, the sum rules give separate quantitative values of the spin and orbital magnetic moments. However, this analysis requires a correct value of the valence hole number and close estimation of the $\langle T_{z}\rangle$ term. For Au nanoparticles, the hole number and $\langle T_{z}\rangle$ values are likely to be quite different from the known values for bulk Au \cite{Zhang}, and hence we did not attempt to estimate these values. We will restrict further discussion to the relative variations of the XMCD signal as a function of external magnetic field or temperature.

Considering the XMCD signal to be proportional to the magnetization, ESM is obtained by recording the peak amplitude of the XMCD spectra at the Au $L_{3}$-edge as a function of external magnetic field and temperature. Figure 3(a) shows ESM measurements scaled arbitrarily for comparison with the magnetic field variation of SQUID magnetization in the same figure. ESM increases with increasing magnetic field without saturation. This behaviour is similar to the magnetization process obtained by SQUID magnetization measurements. As is discussed in the following section, the magnetization process consists of superparamagnetic and temperature independent Pauli-paramagnetic parts. For an ideal superparamagnetic system, the magnetization curve is expressed by $M(H) = N\mu L(x)$, with the Langevin function $L(x) = \coth(x)-1/x$, where $x = \mu H/k_{\rm B}T$, $\mu $ is the magnetic moment per particle and $N$ is the total number of particles per unit mass. Hence, the total magnetization can be expressed as $M(H) = N\mu L(x) + \chi _{\rm Pauli}H$. The dotted line in Fig. 3(a) represents the fitting results. The magnetic moment determined in this way was 0.4 $\mu _{\rm B}$/Au particle. XMCD is an element selective technique and impurity atoms other than Au do not influence the experimental results. Since the value of the magnetic moment is relatively small, one might attribute the magnetization to the polarization of the conduction electrons of Au induced by magnetic impurities. This possibility is denied for the present system by the following reason. First, the amount of magnetic impurity included in the sample was extremely small. We have quantitatively analyzed the amount of the magnetic impurities using an inductively coupled plasma mass spectrometer (ICP-MS), and Fe, Co and Ni, were not detected within the detectable limit of 5 ppm. Even if one assumes that the sample contains Fe impurities of 5 ppm, it is impossible to explain the absolute value of the magnetization measured by SQUID magnetometry. Second, the conduction electrons of Au do not tend to have magnetic polarization induced by the magnetic impurities because the density of states at the Fermi level is small. It is well known that small amounts of magnetic impurities could affect the surrounding magnetic properties significantly and induce ferromagnetic moments for elements on the verge of ferromagnetism, such as Pd and Pt. However, it has not so far been reported that Au exhibits such an effect. Therefore, we conclude that the observed magnetization is an intrinsic effect of the Au particles.

\begin{figure}[tbp]
\includegraphics[width=7cm]{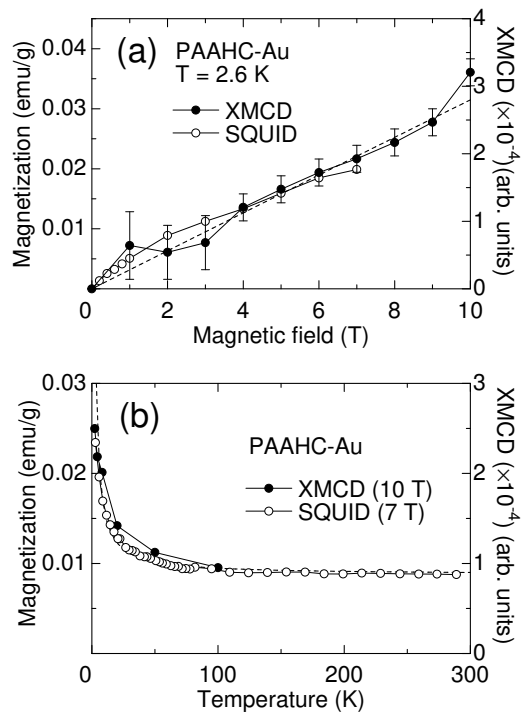}
\caption{\label{ESM} (a) ESM of PAAHC-Au as a function of applied magnetic field and magnetization process obtained by SQUID magnetometer. The integral of the peak intensity yields similar results. Dotted line is the fit to the data assuming a Langevin function plus a linear field-dependent term. (b) Temperature dependence of the XMCD peak intensity at 10 T recorded at the Au $L_{3}$-edge and temperature variation of the magnetization measured by SQUID magnetometer at 7 T. Dotted line is the fit to the data assuming $\chi (T) = N\mu L(x)/H+\chi _{\rm Pauli}$. Solid lines are guides for the eye.}
\end{figure}

The temperature dependence of ESM was also investigated and compared to the temperature variation of the SQUID magnetization. As shown in Fig. 3(b), ESM increases rapidly with decreasing temperature, although it seems that a finite constant value remains at high temperatures. The steep increase of ESM at low temperatures is consistent with the temperature variation of SQUID magnetization and does not contradict the characteristics of the superparamagnetic model (dotted line in Fig. 3(b)). The temperature-independent magnetization corresponds to the Pauli-paramagnetic part mentioned above, which is not observed in bulk gold metal that shows a diamagnetic response (-0.142$\times $10$^{-6}$ emu/g) and monotonic increase with decreasing temperature. We believe that the paramagnetism of Au is masked, in the bulk states, by the huge diamagnetism of the conduction electrons, but not for nanoparticle, the paramagnetism emerges due to a size effect reducing the density of states at the Fermi level, and consequently suppressing the diamagnetism.

Our observation provides evidence of the existence of spin polarization in Au nanoparticles. It should be noted that our results are not related to the induced magnetism by magnetic elements that is often observed in magnetic/nonmagnetic/magnetic layer structures \cite{Wilhelm}, because our system consists of Au and a polymer matrix that is made of nonmagnetic elements (C, H, N). Size effects, such as the so-called Kubo effect, are not responsible for the present results because the Kubo effect should be observed below 1 K for the size of nanoparticles studied here \cite{Volokitin}. One possible explanation for the ferromagnetic polarization is surface ferromagnetism due to the large ratio of the number of atoms on the surface to the number in the core. It has been predicted theoretically that ferromagnetic spin polarization could take place in 4$d$ and 5$d$ transition metals with reduced coordination geometry \cite{Bruegel}. Moreover, if one assumes that superparamgnetism and Pauli-paramagnetism arise from surface atoms and core atoms in Au nanoparticles respectively, the observed mixture of superparamagnetism and Pauli-paramagnetism is reasonably explained.

On the other hand, the polymers, as stabilizers of the nanoparticles, also interact with the surface and would play an important role: the electronic properties of Au particles strongly depends on kind of polymers \cite{Zhang2}. For example, ligand molecules with a large affinity to the metal tend to quench the magnetic moment at the surface of nanoparticles \cite{Leeuwen}. Our recent experiments showed a sizeable reduction in the magnetization for ligand-stabilized nanoparticles such as dodecanthiol-stabilized nanoparticles. On the contrary, several kinds of polymer-stabilized nanoparticles exhibited similar superparamagnetic behavior with large magnetic moments \cite{Hori2}. This suggests that the interaction between the surface of the Au nanoparticles and the ligand is weak for polymer-stabilized nanoparticles compared to dodecanthiol-stabilized nanoparticles, for which it is known that the ligand forms covalent bonds with the metal surface. Therefore, the most reasonable picture for the experimental results is that PAAHC-Au can be regarded as a nearly freestanding cluster so that the surface magnetic moment survives. Supporting this, experimental observations of ferromagnetism or paramagnetism have been reported recently for large bare Pd and Au clusters produced by a gas evaporation method \cite{Taniyama,Li}. For future prospects, variation of the nanoparticle magnetism with the coupling strength between the protective agent and the surface is an interesting subject. The XMCD technique will be a powerful tool for this kind of study.

In summary, our XMCD and ESM experiments have revealed the intrinsic magnetic polarization in the Au nanoparticles with a mean diameter of 1.9 nm. The external magnetic field and temperature dependence of ESM signal suggested that magnetization of Au nanoparticles consist of the superparamagnetic part obeying the Curie law, and the temperature-independent Pauli-paramagnetic part. The mixture of these component is reasonably explained by the picture that the surface atoms are ferromagnetic and the core atoms are Pauli-paramagnetic. We infer that the weak coupling between the protective agent and the surface of the gold is an indispensable condition for the observation of spin polarization.

\begin{acknowledgments}
The synchrotron radiation experiments were performed at SPring-8 with the approval of Japan Synchrotron Radiation Research Institute (JASRI) as Nanotechnology Support Project of The Ministry of Education, Culture, Sports, Science and Technology. (Proposal No. 2002B0380-NS2-np / BL-No.39XU)
\end{acknowledgments}


\end{document}